\documentclass{Interspeech2024}

\interspeechcameraready

\usepackage{hyperref}
\usepackage{url}
\usepackage{graphicx}
\usepackage{booktabs}
\usepackage{multirow}
\usepackage{tikz}
\usepackage[flushleft]{threeparttable}
\usepackage{amsmath}
\usepackage{amssymb}
\usepackage{mathtools}
\usepackage{amsthm}
\usepackage{booktabs}  % For better-looking horizontal rules
\usepackage{siunitx}   % For better number alignment
\usepackage{caption}
\usepackage{multirow}   % For multirow cells
\usepackage{makecell}   % For \multirowcell command
\usepackage[capitalize,noabbrev]{cleveref}
\title{RepCNN: Micro-sized, Mighty Models for Wakeword Detection}
\name{Arnav}{Kundu$^*$}
\name{Prateeth}{Nayak$^*$}
\name{Priyanka}{Padmanabhan}
\name{Devang}{Naik}

\address{
  Apple Inc.}
\email{\{a\_kundu,prateethvnayak,priyanka\_padmanabhan,naik.d\}@apple.com}


\begin{document}
\maketitle
\def\thefootnote{*}\footnotetext{Equal Contribution}\def\thefootnote{\arabic{footnote}}
\begin{abstract}
Always-on machine learning models require a very low memory and compute footprint. Their restricted parameter count limits the model’s capacity to learn, and the effectiveness of the usual training algorithms to find the best parameters. Here we show that a small convolutional model can be better trained by first refactoring its computation into a larger redundant multi-branched architecture.  Then, for inference, we algebraically re-parameterize the trained model into the single-branched form with fewer parameters for a lower memory footprint and compute cost. Using this technique, we show that our always-on wake-word detector model, RepCNN, provides a good trade-off between latency and accuracy during inference. RepCNN re-parameterized models are 43\% more accurate than a uni-branch convolutional model while having the same runtime. RepCNN also meets the accuracy of complex architectures like BC-ResNet, while having 2x lesser peak memory usage and 10x faster runtime.
\end{abstract}
\section{Introduction}
% The design and development of efficient machine learning models
% for always-on applications have advanced significantly,
% witnessing an overall trend of reduced FLOPs (floating-point
% operations) and memory usage while maintaining or improving
% accuracy \cite{HowardMobileNet17,HowardMobileNetv319,kim2021broadcasted}. Often, complex neural
% network architectures \cite{ma2018shufflenet,hu2019squeezeandexcitation,heimdal} are employed to accomplish the balance between the model accuracy and resource demand, but their
% components, such as residual connections make the inference graph complex (although FLOP is low) and lead to increased runtime memory usage and end-to-end inference latency \cite{repvgg, mobileone}.

The design and development of efficient machine learning models for always-on mobile applications have progressed significantly, witnessing an overall trend of reduced FLOPs (floating-point operations) and memory usage while maintaining or improving accuracy \cite{HowardMobileNet17,HowardMobileNetv319,kim2021broadcasted}. This poses a challenge of achieving higher accuracy within limited memory and compute. Specifically related to speech and vision applications on mobile devices, novel architecture optimizations \cite{ma2018shufflenet,hu2019squeezeandexcitation,heimdal} are proposed to achieve high accuracy, but their components, such as residual connections make the inference graph complex (although FLOP is low) and lead to increased runtime memory usage and end-to-end inference latency \cite{repvgg, mobileone}.
% There have been increased focus on maintaining runtime memory usage and latency on edge-processors while designing such architectures\cite{mobileone,repvgg}.
This makes it challenging to deploy these models on resource-constrained devices, ultimately leading to higher latency or unsupported configurations. Due to these strict memory and compute demands, existing models developed for always-on keyword spotting have often been simplistic inference  graphs \cite{sigtia2018efficient,sainath2015convolutional,higuchi2020stacked}.

Although these simpler models are efficient in terms of memory usage, they often fall short in achieving comparable accuracy to more complex multi-branched architectures. One reason for this is that multi-branch topology, like ResNet, create an implicit ensemble of shallower models. This ensemble effect inherently mitigates the gradient vanishing problem, which can be a challenge for single-branch models, allowing more accurate predictions \cite{veit2016residual}.
%Design and development of efficient machine learning models for always on applications have seen a lot of progress with a downward trend in FLOPs and memory usage, while being better in accuracy. However, to develop more accurate models in limited memory the neural network architectures need to be complex \cite{kim2021broadcasted}. But such complex models components like residual connections increase runtime memory usage and latency. This often leads to greater compute demands and runtime memory demands \cite{repvgg, mobileone}, therefore ends up increasing latency on device or not being supported at all. Consequently, adhering to strict memory and compute demands together, the model architectures developed for always on keyword spotting have been simple \cite{sainath2015convolutional,sigtia2018efficient,higuchi2020stacked}. It is challenging for such simple models to reach a comparable level of accuracy as complex multi-branced architectures. An explanation is that a multi-branch topology, e.g., ResNet, makes the model an implicit ensemble of numerous shallower models \cite{veit2016residual}, implicitly avoiding the gradient vanishing problem. %
To this end, we propose a re-parameterizable fully convolutional encoder architecture (RepCNN) that is:
\begin{itemize}
    \item Multi-branched during training to accommodate easier gradient flow.
    \item Simple single branched inference graph to ensure less compute and memory usage during inference execution flow. 
\end{itemize}

% Since, the benefits of multi-branch architectures are mostly restricted to ease of training and add to inference costs we have decoupled the training and inference model graphs. We have also over-parameterized our model during training just like \cite{mobileone} by having multiple kernels of the same size in parallel. These can be fused into a single kernel during inference. Our results indicate such over parameterization helps to achieve better accuracy without having any implications on inference latency.

Considering that the benefits of multi-branch architectures are predominantly seen during training and contribute to increased inference costs, we have proposed a novel approach to decouple the training and inference model graphs for keyword-spotting. By doing so, we can leverage the advantages of multi-branch training while reducing its impact on inference. Further, we hypothesize that using multiple identical kernels in parallel could introduce new learning paths, potentially enhancing the model's accuracy. During inference, these redundant kernels can be combined into a single kernel, without any increase in latency.

\section{Related Works}
Traditional wake-word detection approaches are largely based on Hidden Markov Models which includes acoustic DNN model that computes phoneme probabilities followed usually by \textit{viterbi}-based HMM decoder with output observation scores\cite{sigtia2018efficient,chen2013hybrid,alvarez2019end,prabhavalkar2015automatic,shrivastava2021optimize}, but these systems often suffer from loss-metric mismatches and are difficult to train. With advances in recurrent architectures, there have been complex networks\cite{fernandez2007application,lengerich2016end,hwang2015online} which are easier to train and more accurate, however are not suitable for mobile inference due to high latency concerns. With accuracy gains and decent computational complexities, several End-to-End CNN-based models\cite{HowardMobileNetv319,veit2016residual,ma2018shufflenet,HowardMobileNet17,HowardMobileNetv218} developed for mobile-vision applications provide a great motivation to apply such architectures to wake-word detection. Additionally, convolutional models have the advantage of having streaming audio with input audio length being arbitrary. In this direction, there have been specific single-path CNN-based hand-crafted architectures\cite{sainath2015convolutional,higuchi2020stacked,tang2018deep} that have shown good performance at low latency and runtime memory usage. However, only with complex architectures\cite{kim2021broadcasted,ChoiTCResnet19,li2020smallfootprint,majumdar20interspeech} with high parameter count, bottlenecks or skip-connections we are able to see huge gain in accuracy. Novel residual connections\cite{hu2019squeezeandexcitation} in these work provide further gains, but presence of these residual structures during inference on mobile-applications not only adds computational overhead, but also increase runtime memory size required on-device to store these branched intermediate outputs. Models with such inference graphs become unsuitable when operating in strict memory and computation budget.

Recently, structural re-parameterization has shown significant promise in the field of computer vision with \cite{mobileone,repvgg,chen2022repghost,wang2023repvit}. These models use a multi-branched architecture during training and are converted to a single branched architecture during inference.
Motivated by these methods, we extend the discriminative training work\cite{heimdal,higuchi2020stacked,alvarez2019end} with the approach of multi-branch topology for 1D depth-wise convolutions during training to over-parameterize and increase network capacity, and during inference we linearly combine the convolutional parameters to a trivial architecture that has lesser parameters as well as lower latency and memory usage.
\section{Model Architecture}
As discussed earlier, conventional single branch architectures have poor performance, primarily because of being difficult to train. Multi-branch architectures on the other hand are easier to train while being slower for on-device applications. This is true despite having lower theoretical FLOPs, primarily because of memory access cost \cite{ma2018shufflenet}. Therefore, we use structural re-parameterization to have different training and inference graphs. This section explains our model architecture in detail that incorporates the benefits of multi-branch training graph and converting to a simple inference graph.

\subsection{RepConvBlock}
\label{sec:repconvblock}
\begin{figure}
    \centering
    \includegraphics[width=7.4cm]{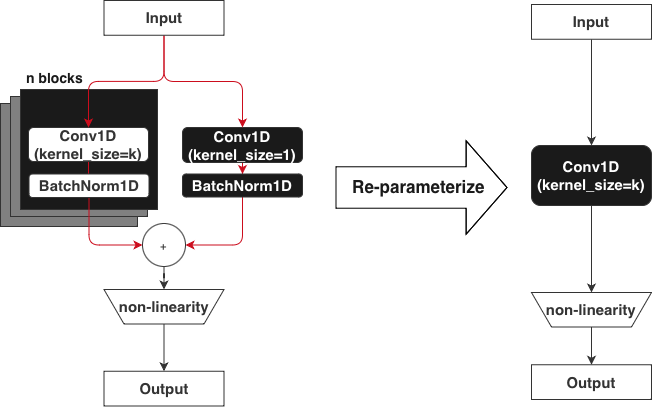}
    \caption{RepConvBlock: Re-parameterizable Convolutional Block}
    \label{fig:repconvblock}
\end{figure}
Based on the principle of structural re-parameterization, we introduce Re-parameterizable Convolutional Block (RepConvBlock) as shown in \cref{fig:repconvblock}. This acts as the foundational element of our work with the training graph shown on the left and inference graph shown on the right. Since we are focusing on tiny ML models for speech processing, the RepConvBlock is designed for 1-D depth-wise separable convolution to be compute and memory efficient \cite{mobileone}. The RepConvBlock consists of 3 architectural elements during training as described below:
\begin{itemize}
    \item Multiple convolutional kernels of the same size \cite{mobileone}, followed by batch normalization layers in parallel in the training graph. This is done to over parameterize the training graph which has proven to be beneficial in prior works \cite{mobileone,repvgg}.
    \item 1x1 convolutional kernel followed by a batch normalization layer parallel to the kernels above to emulate a skip connection.
    \item ReLU layers as non-linearity.
\end{itemize}
During inference the parallel convolutional and batch normalization layers are replaced by an equivalent fused convolutional layer. The fusion process is described in \cref{sec:reparam}. We also employ non-linear layers that are replaced by a clip layer to keep the activations bounded.

\subsection{Reparameterization}\label{sec:reparam}
A 1-D convolutional layer parameterized by $w_k$, followed by a batch normalization layer, during inference can be re-parameterized into another 1-D convolutional layer parameterized by $(W_k,W_b)$ as shown in  \cref{eq:merge_conv_bn}. 
\begin{equation}
\label{eq:merge_conv_bn}
\begin{split}
    y &= bn(\text{Conv}(x))\\
    &= \frac{\alpha_{bn}}{\sigma_{bn}}\times\text{Conv}(x) - \frac{\mu_{bn}}{\sigma_{bn}}+\beta_{bn}\\
    &= \frac{\alpha_{bn}}{\sigma_{bn}}\times w_k * x - \frac{\mu_{bn}}{\sigma_{bn}}+\beta_{bn}\\
    &= W_k * x + W_b
\end{split}
\end{equation}
Here $k$ is the size of the kernel, $W_b$ is the bias of the re-parameterized kernel. $\alpha,\beta,\mu,\sigma$ correspond to the weight, bias, tracked mean and standard deviation for the batch normalization layer, respectively. We re-parameterize individual branches into equivalent convolutional kernels parameterized by $W_k^i$. 

Consequently, parallel convolutional kernels of size $k$ can be fused together into a single kernel by simply adding the corresponding weights. The branch with $\text{kernel size} = 1$ can be represented as a kernel of size $k$ where all weights except the central element are zeros. The re-parameterization of the parallel branches $\{W_1,W_k^1,W_k^2...\}$ is shown in \cref{eq:merge_parallel}. We get a single convolutional kernel ($\hat{W_k}$) to represent the combination of multiple parallel branches as shown in \cref{fig:repconvblock}.
\begin{equation}
\label{eq:merge_parallel}
\begin{split}
    y &= W_1*x + \Sigma_i (W_k^i * x)\\
    &= W_1*x + \Sigma_i (W_k^i) * x \\
    &= \underbrace{[0,\cdots,W_1,\cdots,0]}_{k-elements} + (\Sigma_i W_k^i) * x \\
    &= \hat{W_k} * x
\end{split}
\end{equation}
\begin{figure}
    \centering
    \includegraphics[width=7.4cm]{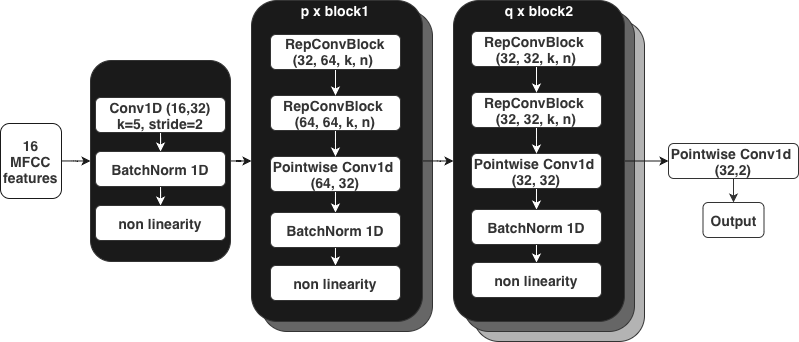}
    \caption{RepCNN training architecture. The \textit{RepConvBlocks} are parameterized as (input-channels, output-channels, kernel-size, num-branches). Here p=q=2 i.e., block1 and block2 are stacked twice. $block1_1$, $block1_2$, $block2_1$, $block2_2$ use kernel-size as $k=7,9,11,13$ respectively. non linearity = ReLU.}
    \label{fig:repcnn}
\end{figure}
\subsection{RepCNN}
The design of the RepCNN model during training is illustrated in \cref{fig:repcnn}. The model comprises of a regular 1D-convolutional layer with kernel size 5 and a stride of 2, followed by a 1D-batch normalization layer and a ReLU layer. The \textit{RepConvBlocks} are used in \textit{block1} and \textit{block2}. In both of these modules we stack 2 \textit{RepConvBlocks} next to one another with the same kernel size followed by a point-wise convolution, batch normalization and ReLU. We use relatively larger kernels in these models with modules of \textit{block1} type using kernels of $k=5$ \& $k=7$, and modules of \textit{block2} type using kernels of $k=11$ \& $k=13$  respectively. We can use these larger kernels without significant increase in runtime latency and memory usage because of absence of branching during inference, which avoid data movement in between nodes of the model as in \cite{mobileone}.

\section{Experiments and Results}
We apply our method for detection of a given wakeword like "Alexa" or "Hey Google" in streaming speech.  The following sections will provide further details on our dataset, training strategies, results, and ablation studies that delve into the impact of various architectural choices made in this paper.

\subsection{Datasets}
We have constructed a robust training dataset of 1M audio utterances, each of which includes the wake-word followed by a user query. These audio samples have been meticulously enhanced to mimic real-world acoustic conditions using room-impulse responses (RIRs) along with echo residuals. To augment the training data and improve model robustness, we have included ambient noise samples from various acoustic environments. Furthermore, we applied a diverse range of training-time gain augmentations, varying from 10dB to -40dB, to simulate various volume levels in real-world scenarios.
For each utterance in training data, we use an acoustic model to get per-frame forced alignments. This provides us with the indices of start and end frame phonemes of the wake-word. We define a positive sample as a window containing the entire wake-word while ending with indices of the end frame. We also harvest random windows with partial or no wake word in them as negative samples. For every positive sample, we pick 20 random negative windows to mimic the real world probability of wake word being present in streaming speech.

The positive test data encompass both near-field and far-field utterances, clearly capturing the wake-word at 3ft and 6ft distances, providing an extensive range of testing scenarios. The negative test data consists of approximately 2000 hours of dense general speech, ensuring a thorough evaluation of the model's ability to distinguish between relevant and irrelevant cues. These datasets have been collected through internal user studies with informed consent approvals.

\subsection{Training and Evaluation}
In order to conduct a comparative study, we trained three models with different architectural designs. The first model is BC-ResNet\cite{kim2021broadcasted}, which serves as an example of a multi-branch architecture and has the same training and inference graph. The second model is a basic stacked 1D convolutional model\cite{higuchi2020stacked} which is used to represent a single branch architecture. The third model is our proposed model, which utilizes stacked \textit{RepConvBlocks} in place of simple 1-D depth-wise convolutional layers during training and is re-parameterized during evaluation to single-branch architecture. All these models have roughly 15k parameters and a receptive field of 1.5 seconds to ensure fair comparison.

% Our training and evaluation methodology aligns with the approach outlined in \cite{heimdal}. 
% We train our model to
% To predict both the end of a keyword and its offset from the start, we equip all our models with a classification head and a regression head.
For input featurization, we apply 26 Mel-Spectral Filter-Banks and extract 16 Mel-frequency cepstral coefficients (MFCCs) at a 10 ms frame rate for every 25 ms of audio, a commonly used approach in speech processing. We use focal-loss \cite{lin2017focal} as the classification loss to handle the skewed distribution of psoitive to negative samples in every batch. To make the best use of limited model capacity, for every batch iteration we sort the loss values of the negative samples and only pick the top-$K$ ($K=50$) hardest negative samples along with all positive sample losses for back-propagation. We also repeat the training with 3 different seeds and the results reported are the average of them.

\begin{table}[!t]
    \centering
    \begin{tabular}{cc}
    \toprule
         Model & FRR (\%) @ 3FA/hr \\
    \midrule
         BC-ResNet\cite{kim2021broadcasted} & 1.66 \\
         Single branch Arch. & 2.88 \\
         \textbf{RepCNN} & \textbf{1.66} \\
    \bottomrule
    \end{tabular}
    \vspace{0.1in}
    \caption{Comparative results of BC-ResNet, single branch architecture, and RepCNN on wake-word detection at 3 FA/hr operating point}
    \label{tab:main_result}
\end{table}
In our evaluation framework, a detected wake-word is considered a true-positive(TP) if it has any overlap with the ground-truth segment. If there is no overlap, it is counted as a false accept (FA). On the other hand, all positive segments containing wake-word where the model fails to detect are classified as false rejects (FR). We measure the False-Reject Rate (FRR) which is proportion of samples containing the wake-word that were rejected from the positive set. And, we measure the number of False-Activation per hour (FA/hr) using the negative set. To compare between different model performances, we report the FRR at a chosen operating point of 3 FA/hr. We also compute the DET-curve characteristic of the different models.
\begin{figure}
    \centering
    \includegraphics[width=7.0cm]{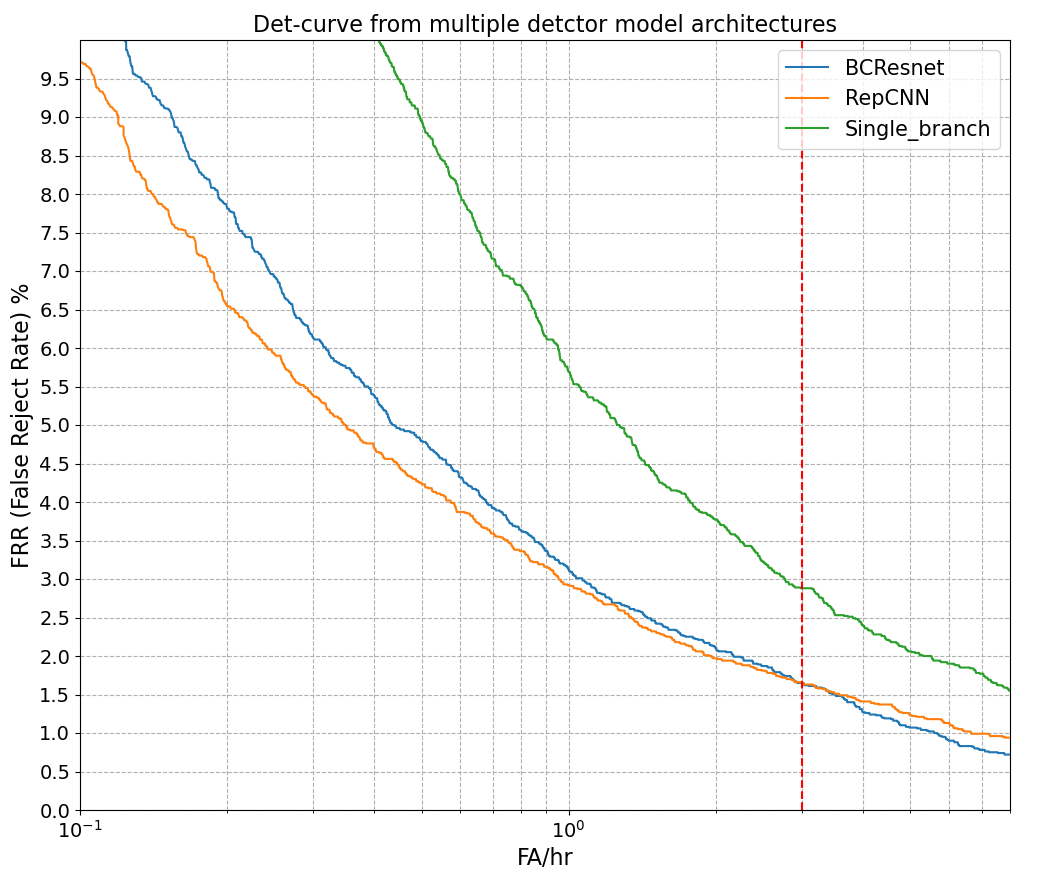}
    \caption{DET curves for model architectures as compared to RepCNN (Orange). We choose 3FA/hr as an ideal operating point for the detection.}
    \label{fig:det_curves}
\end{figure}
\subsection{Results}
For comparison the results for all the candidate models are reported in \cref{tab:main_result}. These results demonstrate that complex multi-branch architecture like BC-ResNet\cite{kim2021broadcasted} are approximately 43\% more accurate than the single branched stacked convolutional model. RepCNN is able to meet or beat this performance accuracy at 3FA/hr operating point or lower, and the DET-curve shown in \cref{fig:det_curves} shows the differences in accuracy at various operating points. We can see that single branch architecture is worse than multi-branch architecture approaches.

% Additionally, we have plotted the DET-curve for the candidate models in \cref{fig:det_curves} to compare the overall performance of the models across a range of FA/hr operating points. On comparing the DET-curves from \cref{fig:det_curves} it can be seen that at lower FA/hr operating points our model (orange) is indeed better than the BC-Resnet model (blue). This signifies that the multi-branch topology with residual blocks greatly helps in training a accurate keyword spotter and at the same time we are able to maintain the prediction accuracy while re-parameterizing it into a single convolutional block.
To compare the overall performance of BC-ResNet vs RepCNN, we compute the ROC-AUC of the models. Both BC-ResNet and RepCNN have 97.9\% AUC score, however BC-ResNet is better than RepCNN at higher FA/hr operating points. To validate that our model has a faster inference speed, we have conducted a on-device latency and memory profile of the BC-ResNet multi-branch architecture against the RepCNN training graph, and the RepCNN re-parameterized model on the CPU of an iPhone Xs. The results are presented in \cref{tab:latency}. The latency results show that the complex inference graph of  RepCNN re-parameterized is 10x faster than BC-ResNet and consumes half the peak runtime memory usage.
\begin{table}[!t]
    \centering
    \small
    \setlength\tabcolsep{3pt} % Adjust column separation
    \begin{tabular}{lS[table-format=1.1]S[table-format=1.1]}%{|p{3cm}|p{1.5cm}|p{3cm}|}
    \toprule
         Model & {Latency (ms)} & {Peak memory (MB)}\\
         % \hline
    \midrule
        BC-ResNet\cite{kim2021broadcasted} &  4.5 & 1.2\\
        RepCNN Training graph (n=2) & 1.8 & 0.8\\
        \textbf{Re-parameterized RepCNN} & \textbf{0.4} & \textbf{0.5}\\
    \bottomrule %\hline
    \end{tabular}
    \vspace{0.1in}
    \caption{Latency and runtime memory usage to produce one output from various model architectures on iPhone Xs CPU.}
    \label{tab:latency}
    \vspace{-0.1in}
\end{table}

This validates our approach that a multi-branch re-parameterizable model can be designed such that it is as accurate as a complicated BC-ResNet model while being more efficient to run on a device for always-on power constrained applications like wake-word triggering. To test its extensibility, we verified that on Google Speech commands \cite{Warden2018SpeechCA}, RepCNN achieves the same accuracy as BCResNet with 50\% faster runtime and 40\% lesser memory.

\subsection{Ablation study}
\begin{figure}
    \centering
    \includegraphics[width=7.8cm]{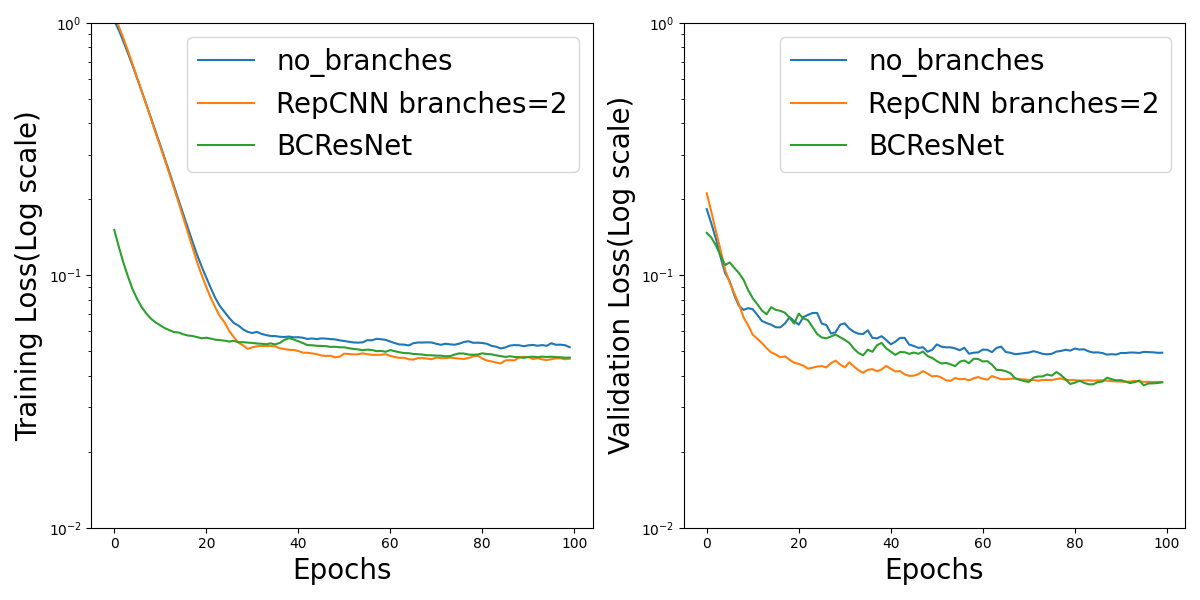}
    \caption{Plot of training and validation loss of RepCNN as compared to a no branch architecture.}
    \label{fig:loss}
\end{figure}

To further validate that our approach is easier to train, we plot the training and validation loss curves of RepCNN in \cref{fig:loss}. For exactly same training setup and hyper-parameters, it can be observed that RepCNN models have a better training and validation loss curve trend compared to the no branch case (a vanilla stacked 1D-CNN similar to \cite{higuchi2020stacked}). This is realized across different seeds of the training as well.
\begin{figure}
    \centering
    \includegraphics[width=7.4cm]{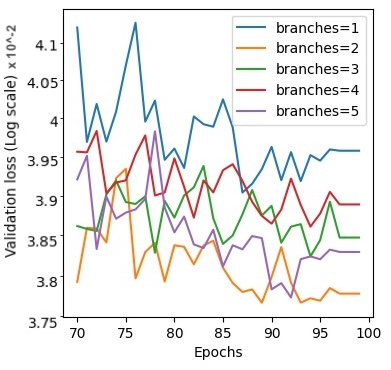}
    \caption{Plot of validation loss of RepCNN with various degrees of over-parameterization. Branches corresponds to the number of parallel 1D depth-wise convolutional kernels apart from the kernel with size 1.}
    \label{fig:loss_zoomed}
\end{figure}

We also explore the effect of different number of branches in the multi-branch topology during training and compare it against the single-branch as part of the ablation study. Primarily, we are interested in comparing the validation loss curves which is an indicator of the training health and also the final performance metrics. In \cref{fig:loss_zoomed}, we verify that over-parameterization does helps to train a better model by looking into the validation loss of various branched versions of RepCNN versus the single-branch architecture. The variations in the loss is due to randomness in sampling hard negative sub-windows and varying augmentations across epochs during training. These models have different degrees of over-parameterization introduced by the number of parallel convolutional kernels (denoted as \textit{n} in \cref{fig:repconvblock}) in RepConvBlock. The final inference graph and parameter count for all these versions is the same after training due to re-parameterization. Additionally, on evaluation we find that using 2 branches gives the best FRR at chosen operating point while using 3,4 or 5 branches is always better than using 1 branch (single-branch) during training as shown in \cref{tab:ablation_k}. It can be observed that allowing multiple branched paths helps to converge to a better model but beyond a limit such redundancy does not tend to improve the metrics.
% Hence, we see that more model capacity helps optimize the model parameters better overall however we also need to prevent any under-fitting through large number of branches.

% DONE? YES ! Uploaded
% Something like however, beyond a limit additional branches are not helpful
\begin{table}[!t]
    \centering
    \begin{tabular}{cccccc}
    \toprule
    Num. branches & 1 & 2 & 3 & 4 & 5\\
    \midrule
    FRR (\%) @ 3 FA/hr & 2.16 & \textbf{1.66} & 1.93 & 1.98 & 2.03\\
    \bottomrule
    \end{tabular}
    \vspace{0.1in}
    \caption{Ablation study on the effect of accuracy with various levels of over-parameterization}
    \label{tab:ablation_k}
    \vspace{-0.2in}
\end{table}
\section{Conclusion}
In this work, we have introduced structural re-parameterization of complex multi-branch models to detect wake words accurately and fast. We have shown that we can take advantage of multi-branch architectures during training without impacting inference latency and runtime memory. We achieve it by designing RepCNN which is a multi-branch architecture during training and is re-parameterized into a single branch architecture during inference. We demonstrate that RepCNN is 10x faster than BC-ResNet. It runs as fast as a regular stacked 1D-CNN while being 43\% better in accuracy. In our further work in this direction, we plan to improve the gap in accuracy with complex architectures, like BC-ResNet at higher FA operating points, through additions like squeeze and excite components while ensuring simplistic inference graph. Furthermore, we aim to explore compression techniques that help in better re-parameterization between multi-branch convolutions.
% The major limitation of this model is it is not as accurate as BCResnet. We believe that some blocks like the squeeze excite block in BCResnet are one of the major contributors to it. In our further investigation we will try to address this gap by looking into these blocks.

% \ifinterspeechfinal
%      The Interspeech 2024 organisers
% \else
%      The authors
% \fi
% would like to thank ISCA and the organising committees of past Interspeech conferences for their help and for kindly providing the previous version of this template.

\bibliographystyle{IEEEtran}
\bibliography{mybib}

\end{document}